\documentclass[prl,aps,twocolumn,superscriptaddress,showpacs]{revtex4}

\usepackage{amsmath}
\usepackage{amssymb}
\usepackage{graphicx}
\usepackage{float}
\usepackage{bbm}
\usepackage{mathptm}

\newcommand{\dx}[0]{{\rm d}x}
\newcommand{\ddx}[0]{{\rm d}^3x}

\newcommand{\rmi}{{\rm i}}
\newcommand{\rme}{{\rm e}}

\newcommand{\ket}[1]{\left| #1 \right\rangle}
\newcommand{\bra}[1]{\left\langle #1 \right|}
\newcommand{\bx}[0]{{\bf x}}

\newcommand{\bm}[0]{{\bf m}}
\newcommand{\bk}[0]{{\bf k}}

\newcommand{\dq}[0]{{\delta \!q}}

\newcommand{\rmm}[0]{{\rm m}}
\newcommand{\be}[0]{{\bf e}}

\begin{document}

\title{Dark state cooling of atoms by superfluid immersion}

\author{A. Griessner}
\affiliation{Institute for Quantum Optics and Quantum Information of the
Austrian Academy of Sciences, A-6020 Innsbruck, Austria} \affiliation{Institute
for Theoretical Physics, University of Innsbruck, A-6020 Innsbruck, Austria}
\author{A. J. Daley}
\affiliation{Institute for Quantum Optics and Quantum Information of the
Austrian Academy of Sciences, A-6020 Innsbruck, Austria} \affiliation{Institute
for Theoretical Physics, University of Innsbruck, A-6020 Innsbruck, Austria}
\author{S. R. Clark}
\affiliation{Clarendon Laboratory, University of Oxford, Parks Road, Oxford OX1
3PU, United Kingdom}
\author{D. Jaksch}
\affiliation{Clarendon Laboratory, University of Oxford, Parks Road, Oxford OX1
3PU, United Kingdom}
\author{P. Zoller}
\affiliation{Institute for Quantum Optics and Quantum Information of the
Austrian Academy of Sciences, A-6020 Innsbruck, Austria} \affiliation{Institute
for Theoretical Physics, University of Innsbruck, A-6020 Innsbruck, Austria}
\date{July 10, 2006}

\begin{abstract}
We propose and analyse a scheme to cool atoms in an optical lattice to
ultra-low temperatures within a Bloch band, and away from commensurate filling.
The protocol is inspired by ideas from dark state laser cooling, but replaces
electronic states with motional levels, and spontaneous emission of photons by
emission of phonons into a Bose-Einstein condensate, in which the lattice is
immersed. In our model, achievable temperatures correspond to a small fraction
of the Bloch band width, and are much lower than the reservoir temperature.
\end{abstract}
 \maketitle

\begin{figure}[tb]
    \begin{center}
        \includegraphics{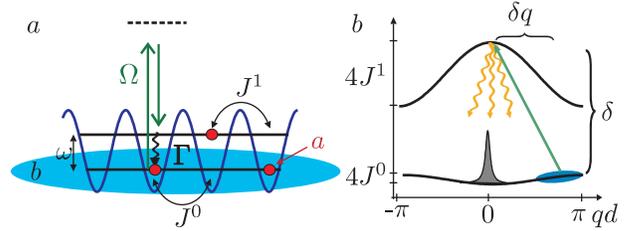}\caption{(a) Cooling setup: Atoms $a$ in an optical lattice
        are coupled to the first excited motional state via a Raman process, and decay to the
ground motional state due to collisional interactions with a BEC of species $b$
in which the lattice is immersed. Tunnelling between neighbouring sites with
amplitude $J^{\alpha}$ gives rise to Bloch bands. (b) Momentum space picture:
Atoms with higher quasi-momentum $q$ are excited to the upper Bloch band, and
decay to random quasi-momentum states. After several cycles, atoms $a$ collect
in a dark state region near $q=0$ with low excitation
probability.}\label{Fig:Setup}
    \end{center}
\end{figure}

Fundamental advances in atomic physics are often linked to the development of
novel cooling methods, as illustrated by laser and evaporative cooling, which
led to the recent realization of degenerate Bose- and Fermi-gases
\cite{BECbook}. This has further led to the achievement of strongly correlated
atomic ensembles in the lowest Bloch band of an optical lattice
\cite{rc_opticallattices1,rc_opticallattices2,rc_olhamiltonians,rc_uli}.
However, in order to realise some of the most interesting condensed matter
phases predicted for lattice Hamiltonians, even better purification of the
motional state is necessary, in particular for atoms in a partially filled
Bloch band \cite{rc_uli}. Here we propose a method for cooling atoms to mean
energies much smaller than the width of the lowest Bloch band $4J^0$. In our
setup (c.f. Fig.~\ref{Fig:Setup}a) lattice atoms $a$ are excited to the first
Bloch band via a Raman laser pulse except when they occupy Bloch states with
quasi-momentum close to zero -- so-called dark states. The lattice is immersed
in a Bose-Einstein Condensate (BEC) of a different atomic species $b$, so that
the atoms can subsequently decay back to the lowest band via collisional
interactions with the BEC reservoir \cite{rc_qubitcooling,rc_fermiloading}.
Thus the atom is recycled to the lowest band by emission of a phonon -- or more
precisely, a Bogoliubov excitation \cite{rc_qubitcooling}. By repeated
application of laser excitation and ``spontaneous emission'', cooling into the
dark state region of quasi-momenta is achieved without loss of atoms
\footnote{This is in contrast to filtering schemes, which work above unit
filling and rely on a gap provided by atomic onsite interactions to achieve
exactly one atom per site \cite{rabl,ignacio_cooling}.} (c.f.
Fig.~\ref{Fig:Setup}b).

This method is inspired by the seminal Kasevich-Chu scheme \cite{raman_chu} for
sub-recoil laser cooling \cite{lasercooling,Levy}, but replaces internal atomic
states by Bloch band excitations, and spontaneously emitted photons by phonons.
Our scheme thus operates on a much smaller energy scale than laser cooling,
with correspondingly lower temperatures. This method can also be seen as a form
of sympathetic cooling, where the energy is removed by phonons with energies
equal to the Bloch band separation. Such phonon modes will initially be in the
vacuum state, giving an effective $T=0$ reservoir, and allowing temperatures
\textit{significantly lower than the BEC reservoir temperature}, in contrast to
standard sympathetic cooling. The ability to switch the collisional
interactions via Feshbach resonances \cite{rc_feshbach} enables us to study the
cooling scenario in the weakly interacting gas, creating strongly correlated
phases by ramping up the interaction in a final step.

On a formal level our cooling scheme can be written as the iterative
application of a map,
\begin{align}
    \mathcal{M}_j:\hat \rho_{j} \rightarrow \hat \rho_{j+1}\equiv\left({\hat{\mathcal D}}
    \circ\hat{E}_{j}\right)\hat \rho_{j},\label{map}
\end{align}
where the density operator $\hat \rho_j$ describes the atoms in the lowest
Bloch band before the $j$th step. Each step consists of two parts: the coherent
laser excitation $\hat E_j$ of lattice atoms $a$, and the dissipative decay
$\hat D$ returning atoms to the lowest band via coupling to the reservoir $b$
(Fig.~\ref{Fig:Setup}). To achieve the best possible cooling, differently
shaped excitation pulses $\hat{E}_{j}$, $j=0,\dots,N_p-1$ are applied, and this
sequence is repeated, with $\hat{E}_{j}=\hat{E}_{j \;{\rm mod} \;N_p}$. This
repeated application of the map corresponds to a purification of the density
operator, starting from an initial mixed state (e.g., a thermal distribution)
towards a pure state (at zero temperature, $T=0$). In order to find appropriate
forms of the Raman pulses and the action of $\hat D$, we analyse the dynamics
of the lattice atoms and their interaction with the reservoir gas.

We consider a one dimensional model for the motion of atoms $a$, which is
readily generalised to higher dimensions. Including Raman coupling, the
Hamiltonian is $\hat H_a= \hat H_0+\hat H_I$, with
\begin{align}
      \hat H_0&=&\sum_{q,\alpha} \varepsilon_q^\alpha \left(\hat A_{q}^{\alpha}\right)^\dagger
      \hat A_{q}^\alpha+\left(\omega-\delta\right)\sum_q \left(\hat A^1_{q}\right)^\dagger \hat A^1_{q}\nonumber\\
      & &+\frac{\Omega}{2}\sum_q\left[ \left(\hat A^1_{q}\right)^\dagger \hat A^0_{q-\dq}
      +{\rm h.c.}\right],\\
      \hat H_I&=&\frac{1}{2}\sum_{i,\alpha} U^{\alpha\alpha} \hat{n}^\alpha_{i}(\hat{n}^\alpha_{i}-1)+U^{10}\sum_{i} \hat{n}^1_{i}\hat{n}^0_{i}. \label{Hai}
\end{align}
Here, $\hat A^\alpha_{q}$ and $(\hat A^\alpha_{q})^\dagger$ are annihilation
and creation operators for quasi-momentum $q$ in Bloch band $\alpha \in \{0,1
\}$, satisfying Bose or Fermi (anti-)commutation relations. The kinetic energy
is $\varepsilon^\alpha_q=-2J^\alpha \cos(qd)$, where $d$ is the lattice
spacing, $J^\alpha$ are the tunnelling amplitudes (with $J^0>0$ and $J^1<0$,
see Fig.~\ref{Fig:Setup}b), and $\omega$ is the band separation. The effective
Rabi frequency $\Omega=\Omega_R\int \dx\, {\rm exp}(-\rmi \dq x)w^1(x)w^0(x)$,
where $\Omega_R$ is the two photon Rabi frequency as a function of time during
the pulse and $w^\alpha(x)$ the Wannier functions. The Hamiltonian is a
two-band model, written in a rotating frame with Raman detuning $\delta$, and
$\dq$ denotes the momentum transfer. We choose units $\hbar=k_B=1$, where $k_B$
is the Boltzmann constant. The parameters $\Omega$, $\delta$ and $\dq$ will
change during the pulse sequence, but be constant during a given pulse $j$.
Onsite interactions between lattice atoms are represented by $\hat H_I$, with
$\hat{n}^\alpha_{i}$ the number operator for atoms in site $i$ and band
$\alpha$, and $U^{\alpha\alpha'}$ the associated onsite energy shifts
\cite{rc_opticallattices1}.

The density-density interaction between lattice atoms $a$ and a three
dimensional BEC reservoir $b$, which gives rise to the decay $\hat{\mathcal
D}$, is described by the Hamiltonian \cite{rc_qubitcooling}
\begin{align}
  \hat  H_{\rm int}=\sum_{\alpha,\alpha'}\sum_{\bk, q}\left(G_{\alpha,\alpha'}^\bk \hat b_\bk
  \left(\hat A^\alpha_{q}\right)^\dagger
   \hat A^{\alpha'}_{q-k}+{\rm h.c.} \right).\label{Hint}
\end{align}
Here, the operator $\hat b_\bk^\dagger$ creates a Bogoliubov excitation with
momentum $\bk=(k,k_y,k_z)$, and neglecting overlap of Wannier functions in
different lattice sites, the coupling $G_{\alpha,\alpha'}^{\bk}\approx
g_{ab}(S(\bk)\rho_b /V)^{1/2} \int \ddx \rme^{\rmi\bk\bx} w^\alpha(\bx)
w^{\alpha'}(\bx)$. The strength of the inter-species contact interaction is
denoted $g_{ab}$, $\rho_b$ and $V$ are the density and volume of the BEC
reservoir, and $S(\bk)$ is the static structure factor \cite{BECbook}. For
excitation energies less than the chemical potential $\mu$, excitations are
sound waves for which $S(\bk)$ is strongly suppressed, and $S(\bk)\rightarrow
0$ as $|\bk|\rightarrow 0$ \cite{BECbook}. For energies larger than $\mu$,
excitations are in the particle-like sector of the spectrum, with much larger
$S(\bk) \rightarrow 1$. Here we will typically have $4J^0 < \mu < \omega$, so
that decay between bands is induced by particle-like excitations with strong
coupling, but collisional processes between the reservoir and atoms in the
lowest Bloch band are suppressed. In close analogy to
Ref.~\cite{rc_qubitcooling,rc_fermiloading} we derive a master equation for the
reduced system density operator, describing the decay between bands in the
Born-Markov approximation \footnote{We neglect effects due to reabsorption,
which are discussed in the context of laser cooling \cite{reabsorption}. These
effects can be suppressed, e.g., by evaporative cooling of the reservoir
\cite{rc_fermiloading}.}. The associated Liouvillian is
$\mathcal{L[\rho]}=\sum_k \Gamma_k \left( 2c_k\rho c_k^\dagger-c_k^\dagger c_k
\rho-\rho c_k^\dagger c_k\right)/2$. The momentum $k$ along the lattice axis is
bounded by $|k|\leq \sqrt{2m_b\omega}$ due to energy conservation, where $m_b$
is the mass of atoms $b$, and the jump operators $c_k$ are defined as
$c_k=\sum_q A_{q-k,0}^\dagger A_{q,1}$. The resulting decay rates $\Gamma_k$
for spontaneous emission of a phonon with momentum $k$ projected on the axis of
the lattice can be written explicitly for deep lattices, where $\omega\gg
|J^1|,J^0$ and the individual lattice sites can be approximated as harmonic
oscillator potentials. We find $\Gamma_k= g_{ab}^2\rho_bm_a a_0^2k^2
\exp(-a_0^2k^2/2)/2L$, with $a_0$ the size of the ground state in each lattice
site, $m_a$ the mass of atoms $a$, and $L$ the length of the 1D lattice. We
denote the total decay rate from the excited band by $\Gamma=\sum_k \Gamma_k$.
We consider a situation where dissipation is switched off during the excitation
step, so that $\hat E_j$ and $\hat D$ occur separately. This can be achieved
e.g., by tuning the collisional interaction so that $g_{ab}\approx 0$. We can
read the action of $\hat E_j$ and $\hat D$ for a given step $j$ from the master
equation.

We first illustrate the cooling process for a single lattice atom, designing a
sequence of Raman laser pulses, where the $j$-th pulse excites the atom with
initial quasi-momentum $q$ in the lowest band to the first excited band with
probability $P_j(q)$. We require $P_j(q)=0$ for $q\approx 0$, but
$P_j(q)\rightarrow 1$ for states with high quasi-momentum (c.f.
Fig.~\ref{Fig:Setup}b). In analogy with Raman cooling schemes in free space
\cite{raman_chu} we choose square pulses with duration $\tau_j=\pi/\Omega_j$,
for which
$P_j(q)=\sin^2(\sqrt{\delta_{q+\dq_{\!j}}^2+\Omega_j^2}\tau_j/2)\Omega_j^2/(\delta_{q+\dq_{\!j}}^2+
\Omega_j^2)$, with the effective detuning
$\delta_{q+\dq_{\!j}}\equiv\omega+\varepsilon^1_{q+\dq_{\!j}}-\varepsilon^0_q-\delta_j$.

\begin{center}
   \begin{figure}[tb]
       \includegraphics{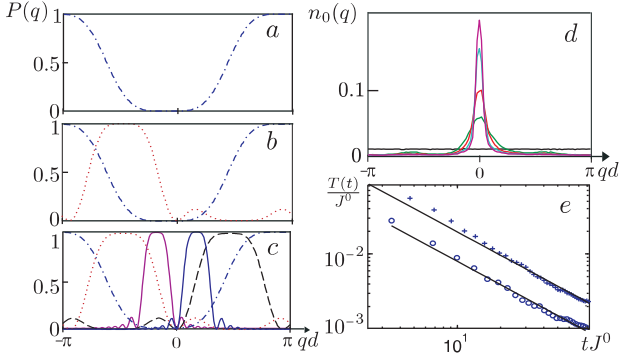}\caption{(a)-(c) Excitation probability
$P -j(q)$
       for a sequence of $N_p=5$ pulses: first (a-c, dash-dot),
   second  (b-c, dotted), and the remaining pulses (c). Parameters
       used: $\Omega=(27.9,13.7,13.7,2.37,2.37)J^0$,
       $\dq=(0.16,-1.75,1.75,-2.63,2.63)/d$,
       $(\delta-\omega)=-(28.4,25.8,25.8,24.7,24.7)J^0$ for the different pulses.
       (d) Successive narrowing of the momentum distribution in the lower
       Bloch band after $0,1,3,5$ and $10$ cooling cycles from numerical simulations ($M=101$ lattice sites)
        based on the pulse sequence in (a)-(c).
       We choose parameters for $^{87}$Rb in the lattice
       and $^{23}$Na in the reservoir, with $\Gamma=53J^0$ from $m_a/m_b=3.73$,
       $\rho_b=5\times 10^{14}{\rm cm}^{-3}$ and scattering length $a_{ab}=14$nm.
       $V_0=10\omega_R$, and $\omega_R=2\pi\times 3.8$kHz.
       (e) Temperature vs.~time for a single atom: crosses and circles denote numerical,
       solid lines analytical results based on L\'evy statistics.
       Pulse sequences for circles: same as in (d); crosses: $N_p=3$ pulses with
       $\Omega=(32.6,7.9,7.9)$, $\dq d=(0.31,2.12,-2.12)$, $(\delta-\omega)=
       -(28.4,25.3,25.3)J^0$.}\label{Fig:pulses}
   \end{figure}
\end{center}

\begin{center}
   \begin{figure}[tb]
       \includegraphics{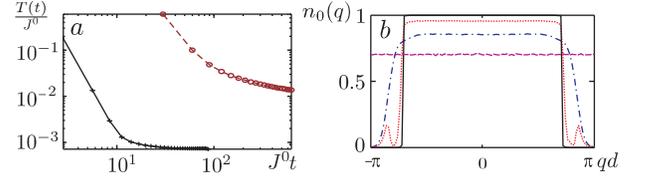}\caption{
       Numerical simulation of the QBME:
       (a) Temperature as a function of cooling time for bosons (crosses) and fermions (circles).
       (b) Momentum distribution in band $\alpha=0$ for fermions after 0 (dashed),
       1 (dash-dot), 2 (dotted), and 20 (solid) cooling cycles, each with $N_p=4$ pulses.
       Parameters used: Bosons: As for Fig.\protect{\ref{Fig:pulses}}a-c, but
       $N=51$ particles.
       Fermions: $N=71$, $M=101$, Blackman pulses with $\tau J^0=(1.78,1.78,6.8,6.8)$,
       $\dq d=(0.19,-0.19,0.75,-0.75)$, $(\omega-\delta)=(28.4,28.4,27.9,27.9)J^0$; $V_0=10\omega_R$,
       $\omega_R=2\pi\times 6$kHz, $m_a/m_b=1.74$ and
       $\Gamma=52.6J^0$.}\label{Fig:manybodypulses}
   \end{figure}
\end{center}

\begin{center}
   \begin{figure}[tb]
       \includegraphics{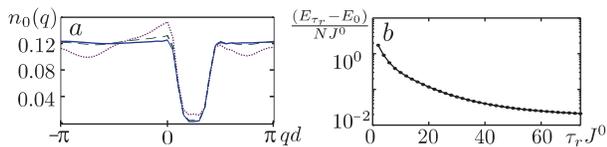}\caption{(a) Momentum distribution in the lowest
       Bloch band from numerical simulations after a single excitation pulse beginning with equal
       occupation, for interaction strengths $U^{\alpha \alpha'}=(0,0.1,0.3)J^0$ (solid, dashed, and dotted lines).
       Parameters used: $N=5$ atoms in $M=41$ lattice sites $\delta q d =1$, $\Omega=1.05J^0$,
       $(\delta-\omega)=27.9J^0$, $V0=10\omega_R$, $\omega_R=2\pi\times 3.8$kHz.
       (b) Energy $E_{\tau_r}$ of the state obtained by beginning in the $U^{00}=0$ ground
       state with $N=10$, $M=21$, and ramping the interaction strength as from $U^{00}\approx 0$ to
       $U^{00}\approx 20J^0$ in time $\tau_r$, as $U^{00}=20J^{0}(1 - \{1 + \exp[(t - \tau_r/2)/(\tau_r/10)]\}^{-1})$.}\label{adiabatic}
   \end{figure}
\end{center}

An example of an efficient pulse sequence is shown in Fig.~\ref{Fig:pulses}a-c.
We begin with an intense laser pulse which resonantly excites atoms with
momentum $q\sim \pi/d$ around the edges of the Brillouin zone
(Fig.~\ref{Fig:pulses}a). The subsequent pulses move the resonant transition
closer to $q=0$ by adjusting the momentum transfer $\dq_{\!j}$ and Raman
detuning $\delta_j$ (Fig.~\ref{Fig:pulses}b). In order to prevent excitation of
atoms with $q=0$, we decrease $\Omega$ and increase the pulse duration, $\tau$
for the later pulses, each time achieving $P_j(q=0)=0$. To resolve the
transition we must always have $\Omega\ll 8|J^1|$, and therefore $\tau\gg
\pi/8|J^1|$. Note that it is the value of $J^1$ and not $J^0$ that gives the
resolution of the excitation. However, the relationship between $J^1$ and $J^0$
is fixed by the lattice depth (e.g., for the parameters used in
Figs.~\ref{Fig:pulses}--\ref{adiabatic}, $V_0=10\omega_R$, we have
$J^0=0.019\omega_R$ and $J^1=-0.25\omega_R$). By combining a sequence of 5
pulses, one can efficiently excite most atoms with $|q|>0$, as shown in
Fig.~\ref{Fig:pulses}c. Note that the pulse sequence has been carefully
designed to avoid significant population outside the lowest two bands.

To quantitatively analyze the cooling process we numerically simulate the time
evolution of the density operator $\hat \rho$ using quantum trajectories
\cite{MCWF} starting from a completely mixed state in the lowest band ($T\gg
4J^0$), with  $\sim 10^5$ trajectories. During the cooling process, the
momentum distribution develops a sharp peak near $q=0$ after very few
iterations as shown in Fig.~\ref{Fig:pulses}d. In Fig.~\ref{Fig:pulses}e we
plot the temperature of the state as defined by $k_BT/2=\pi^2\sin^2(\Delta
\!q\,d)(J^0)^2/\omega_R$, where $\Delta\!q$ is the half-width of the momentum
distribution at $e^{-1}$ of the maximum value. We find excellent agreement
between our numerical calculations and analytical results obtained with L\'evy
statistics \cite{raman_chu,Levy} as a function of time. The latter predicts a
final temperature $T\propto t^{-1}$ for square pulses, as shown in
Fig.~\ref{Fig:pulses}e. For a zero-temperature reservoir, and parameters as in
the caption for Fig.~\ref{Fig:pulses}, we reach temperatures $T\sim 2\times
10^{-3}J^0$ in time $t_fJ^0\sim 30$.

Finite temperature $T_b$ in the reservoir can lead to sympathetic heating of
lattice atoms $a$ by absorption of thermal phonons, as described by $\hat
H_{\rm int}$. However, this process is forbidden by energy and momentum
conservation, provided $J^0 < \sqrt{\mu \omega_R m_a /(2 m_b)}/\pi$. In detail,
energy conservation requires $c |\bk| =\epsilon^0_q - \epsilon^0_{q'}$ and
conservation of momentum along the lattice direction leads to $k=q-q'$ ($|k|
\leq |\bk|$), where the atom $a$ is scattered from quasi-momentum $q\approx 0
\rightarrow q'$ by absorption of a phonon with momentum $\bk$, and $c$ is the
sound velocity in the BEC. These conditions cannot be fulfilled unless the
above inequality is violated. Higher order processes involving two or more
thermal phonons will be small. In numerical simulations we also checked that
the cooling protocol is insensitive to small timing errors. Whilst in the above
protocol we have switched off the decay during application of $\hat E_j$, we
can leave decay switched on, provided that $\Gamma \ll 1/ \tau \ll |J^1|$. This
will restrict the length of the pulses that can be applied, thus slowing the
cooling process.

The cooling scheme can be readily adapted to many bosons or fermions. For
bosons, we assume that the collisional interaction between atoms $a$ is tuned
to zero ($\hat H_I\rightarrow 0$). We work out the efficiency of the cooling
protocol by deriving a quantum Boltzmann master equation (QBME) \cite{QKT},
which describes transitions between classical configurations of atoms occupying
momentum states in the Bloch bands, ${\bf m}=[
\{\rmm^{0}_{q}\}_{q},\{\rmm^{1}_{q}\}_{q}]$, where $\rmm_q^\alpha$ is the
occupation number of quasi-momentum state $q$ in band $\alpha$. This is derived
from the master equation by projection of the density operator $\rho$ onto
diagonal elements, $\mathcal{P}\hat \rho\mathcal{P}=\sum_{\bf m} w_{\bf
m}\ket{\bf m}\bra{\bf m}$, neglecting off-diagonal coherences. For the
excitation step, $\hat E_j$, the evolution is computed exactly from the
excitation probability $P_j(q)$, and for the decay, $\hat D$, we obtain
\begin{align}
    \dot{w}_{\bm}=&\sum_{k,q}\Gamma_k\left[ \rmm_{q-k}^0(1\pm\rmm_{q}^1)w_{\bm'}
    - \rmm_{q}^1 (1\pm\rmm_{q-k}^0)w_{\bm}\right],\nonumber%\label{QBME}
\end{align}
where $\bm'=\bm-\be_{q-k,q}$ is the resulting configuration when a particle
with quasi-momentum $q$ in the upper band decays to the lower band with new
quasi-momentum $q-k$, i.e., $\be_{q-k,q}$ is a configuration vector with
$\rmm_{q-k}^0=1$, $\rmm_{q}^1=-1$ and all other entries zero. The upper (lower)
signs are for bosons (fermions). The approximation inherent in neglecting
off-diagonal coherences only plays a role during the decay step, where these
coherences couple to the occupation probabilities. We remark that an
\textit{exact} physical realisation of the QBME can be obtained by modulating
the lattice depth after each excitation step, randomising the off-diagonal
elements \cite{twirl}.

Fig.~\ref{Fig:manybodypulses}a shows the decrease in temperature as a function
of time for bosons and fermions, obtained from monte carlo simulations of the
QBME \cite{QKT}. For bosons, we use the same excitation pulse sequence as for a
single atom in Fig.~\ref{Fig:pulses}. The cooling process in this case
outperforms that for a single atom, reaching low temperatures on shorter times
due to bosonic enhancement (here we compute temperature as for a single atom,
but fitting a Gaussian to the $q\neq 0$ momentum distribution). For fermions,
the pulse sequence must be changed to create a dark state region of
quasi-momenta with $|q|<q_F$, where $q_F$ is the Fermi momentum, in order to
cool towards a $T=0$ Fermi distribution. In this case, time-square pulses are
no longer efficient as there is a large secondary peak in $P_j(q)$ (see
Fig.~\ref{Fig:pulses}b), and we instead use Blackman pulses \cite{raman_chu},
which approach $P_j(q)=0$ monotonically. The momentum distribution develops the
expected shape of a cold Fermi distribution after very few iterations
(Fig.~\ref{Fig:manybodypulses}b) (we compute temperatures for fermions by
fitting a Fermi distribution to these results).

Our model predicts that the temperature will always decrease with increasing
cooling time. Experimental imperfections will, in practice, give rise to
decoherence and heating (e.g., from spontaneous emissions
\cite{rc_opticallattices1}, collisions with background gases, scattering
multiple phonons). One assumption made above for bosons was that the
interaction between lattice atoms $a$ is approximately zero. Using time
dependent DMRG methods \cite{vidal} we computed the population remaining in the
lowest band after an excitation pulse with interactions present
(Fig.~\ref{adiabatic}a). For $U^{\alpha \alpha'} \ll 1/ \tau \ll |J^1|$ there
is no significant change in the excitation profile and the above conclusions
should not change.

Finally, after the cooling we adiabatically ramp up the interaction strength to
produce an interacting system, assuming that the system is decoupled from the
reservoir $b$. This is illustrated in Fig.~\ref{adiabatic} for the case of
ramping from a non-interacting to a hard-core Bose lattice gas in 1D (Tonks
gas). We compute the evolution from a Bose-Hubbard model ($H_a$ with no Raman
coupling, and only the lowest band), again using time dependent DMRG methods
\cite{vidal}. The energy deposited during the ramp of the interaction strength
is plotted as a function of the ramp time $\tau_r$, and for ramping times on
the order of $10/J^0$, we observe negligible heating within the Bloch band.
Numerical tests using example excited states also showed that the energy
difference between different initial states is not significantly increased
during the ramp.

In summary, filtering quasi-momentum states in the lowest Bloch band
of an optical lattice and recycling them by ``spontaneous emission''
of phonons combine to give a cooling scheme producing temperatures a
small fraction of the lowest Bloch band width. These temperatures
are a necessary step towards the realisation of strongly correlated
quantum phases not currently accessible in optical lattice
experiments.

We thank P. Rabl for discussions. This work was supported by OLAQUI.
Work in Innsbruck was supported by Austrian Science Foundation,
SCALA, and IQI; and  work in Oxford by
 EPSRC project EP/C51933/1.


\begin{thebibliography}{99}
\bibitem{BECbook}L. Pitaevskii and S.
Stringari, \textit{Bose-Einstein Condensation} (Oxford University Press,
Oxford, 2003).
\bibitem{rc_opticallattices1}D. Jaksch and P. Zoller, Annals of Physics \textbf{315}, 52
(2005); M. Lewenstein \textit{et al.}, cond-mat/0606771.
\bibitem{rc_opticallattices2} K. Winkler \textit{et al.}, Nature \textbf{441}, 853 (2006); S. F\"olling \textit{et al.},
cond-mat/0606592; G. K. Campbell \textit{et al.}, cond-mat/0606642; T. Volz
\textit{et al.}, cond-mat/0605184; J. Sebby-Strabley \textit{et al.},
cond-mat/0602103; K. G\"unter \textit{et al.}, Phys. Rev. Lett. \textbf{96},
180402 (2006); S. Ospelkaus \textit{et al.}, \textit{ibid.} \textbf{96}, 180403
(2006); L. Fallani \textit{et al.}, cond-mat/0603655.
\bibitem{rc_olhamiltonians}A. Micheli, G. K. Brennen and P. Zoller, Nature Physics \textbf{2}, 341 (2006);
L. Santos \textit{et al.}, Phys. Rev. Lett. {\bf 93} 030601 (2004); J. J.
Garcia-Ripoll, M. A Martin-Delgado, and J. I Cirac, \textit{ibid.} \textbf{93},
250405 (2004).
\bibitem{rc_uli}S. Trebst, U. Schollw\"{o}ck, M. Troyer, and P. Zoller, \textit{ibid.} \textbf{96}, 250402 (2006).
\bibitem{rc_qubitcooling} A. J. Daley, P. O. Fedichev, and P. Zoller Phys. Rev. A {\bf 69}, 022306 (2004).
\bibitem{rc_fermiloading} A. Griessner, A. J. Daley, D. Jaksch, and P. Zoller, Phys. Rev. A \textbf{72}, 032332 (2005).
\bibitem{raman_chu}M. Kasevich and S. Chu, Phys. Rev. Lett. \textbf{69}, 1741 (1992); J. Reichel \textit{et al.}, \textit{ibid.} \textbf{75}, 4575 (1995).
\bibitem{lasercooling}H. J. Metcalf and P. van der Straten, \textit{Laser cooling and trapping}, (Springer-Verlag, New York,
1999).
\bibitem{Levy} F. Bardou, J. P. Bouchaud, A. Aspect and C. Cohen-Tannoudji, \textit{L\'evy Statistics and Laser Cooling}, (Cambridge University Press, Cambridge, 2002).
\bibitem{rc_feshbach}E. A. Donley, N. R. Claussen, S. T. Thompson, and C. E. Wieman, Nature
        \textbf{417}, 529 (2002); T. Loftus \textit{et al.}, Phys. Rev. Lett. \textbf{88}, 173201 (2002); M. Theis \textit{et al.}, \textit{ibid.}, \textbf{93}, 123001 (2004).
\bibitem{MCWF} See C. W. Gardiner and P. Zoller, \textit{Quantum Noise}, Third Edition (Springer, Berlin, 2005) and references therein.
\bibitem{QKT} C. W. Gardiner and P. Zoller Phys. Rev. A \textbf{55}, 2902 (1997); D. Jaksch, C. W. Gardiner, and P. Zoller, Phys. Rev. A \textbf{56}, 575 (1997).
\bibitem{twirl} This is
reminiscent of the twirl in state purification protocols: D. Deutsch
\textit{et al.}, Phys. Rev. Lett. \textbf{77}, 2818 (1996).
\bibitem{vidal}G. Vidal, Phys. Rev. Lett. \textbf{91}, 147902 (2003); A. J. Daley \textit{et al.}, J. Stat. Mech.: Theor. Exp. P04005 (2004); S.R. White and A.E. Feiguin, Phys. Rev. Lett. \textbf{93},
076401 (2004); F. Verstraete, J. J. Garcia-Ripoll, and J. I. Cirac,
\textit{ibid.} \textbf{93}, 207204 (2004).
\bibitem{rabl}P. Rabl, \textit{et al.}, Phys. Rev. Lett. \textbf{91}, 110403 (2003).
\bibitem{ignacio_cooling}M. Popp, J. J. Garcia-Ripoll, K. G. H. Vollbrecht, and J. I. Cirac,
cond-mat/0603859.
\bibitem{reabsorption}Y. Castin, J. I. Cirac, and M. Lewenstein, Phys. Rev. Lett. \textbf{80}, 005305
(1998).
\end{thebibliography}
\end{document}